# Electrical and structural characterization of *in-situ* MOCVD $Al_2O_3$/β-$Ga_2O_3$ and $Al_2O_3$/β-$(Al_xGa_{1-x})_2O_3$ MOSCAPs

A F M Anhar Uddin Bhuiyan[1, 3], Lingyu Meng[1], Dong Su Yu[1], Sushovan Dhara[1], Hsien-Lien Huang[2], Vijay Gopal Thirupakuzi Vangipuram[1], Jinwoo Hwang[2], Siddharth Rajan[1, 2] and Hongping Zhao[1, 2, a)]

[1]*Department of Electrical and Computer Engineering, The Ohio State University, Columbus, OH 43210, USA*

[2]*Department of Materials Science and Engineering, The Ohio State University, Columbus, OH 43210, USA*

[3]*Department of Electrical and Computer Engineering, University of Massachusetts Lowell, Lowell, MA 01854, USA*

[a)]Corresponding author Email: zhao.2592@osu.edu

## Abstract

This study investigates the electrical and structural properties of metal-oxide-semiconductor capacitors (MOSCAPs) with *in-situ* metal-organic chemical vapor deposition (MOCVD)-grown $Al_2O_3$ dielectrics deposited at varying temperatures on (010) β-$Ga_2O_3$ and β-$(Al_xGa_{1-x})_2O_3$ films with different Al compositions. The $Al_2O_3$/β-$Ga_2O_3$ MOSCAPs exhibited a strong dependence of electrical properties on $Al_2O_3$ deposition temperature. At 900°C, reduced voltage hysteresis (~0.3V) with improved reverse breakdown voltage (74.5V) were observed, corresponding to breakdown fields of 5.01 MV/cm in $Al_2O_3$ and 4.11 MV/cm in β-$Ga_2O_3$ under reverse bias. In contrast, 650°C deposition temperature resulted in higher voltage hysteresis (~3.44V) and lower reverse breakdown voltage (38.8V) with breakdown fields of 3.69 and 2.87 MV/cm in $Al_2O_3$ and β-$Ga_2O_3$, respectively, but exhibited impressive forward breakdown field, increasing from 5.62 MV/cm at 900°C to 7.25 MV/cm at 650°C. High-resolution scanning transmission electron microscopy (STEM) revealed improved crystallinity and sharper interfaces at 900 °C, contributing to enhanced reverse breakdown performance. For $Al_2O_3$/β-$(Al_xGa_{1-x})_2O_3$ MOSCAPs, increasing Al composition (x) from 5.5% to 9.2% reduced net carrier concentration and improved reverse breakdown field contributions from 2.55 to 2.90 MV/cm in β-$(Al_xGa_{1-x})_2O_3$ and 2.41 to 3.13

MV/cm in $Al_2O_3$. The electric field in $Al_2O_3$ dielectric under forward bias breakdown also improved from 5.0 to 5.4 MV/cm as Al composition increases from 5.5% to 9.2%. The STEM imaging confirmed the compositional homogeneity and excellent stoichiometry of both $Al_2O_3$ and β-$(Al_xGa_{1-x})_2O_3$ layers. These findings demonstrate the robust electrical performance, high breakdown fields, and excellent structural quality of $Al_2O_3$/β-$Ga_2O_3$ and $Al_2O_3$/β-$(Al_xGa_{1-x})_2O_3$ MOSCAPs, highlighting their potential for high-power electronic applications.

## I. Introduction

Ultrawide bandgap β-$Ga_2O_3$ semiconductor has attracted significant attention for its potential in high power device applications due to its excellent properties, including its ultrawide bandgap energy (~4.85 eV), predicted high breakdown field strength (8 MV/cm), controllable n-type doping capability, and bandgap engineering by alloying with $Al_2O_3$ [1-31]. To fully exploit its capabilities in high-performance power devices, the integration of compatible dielectric materials with properties such as a substantial conduction band offset, the ability to endure high electric fields, and a high-quality interface with the semiconductor characterized by low interface and bulk defect density is essential. Various insulators, including $SiO_2$, $Al_2O_3$, SiN, $HfO_2$, and high-permittivity materials such as $BaTiO_3$, have been investigated for their potential as gate oxide materials and passivation layers in β-$Ga_2O_3$ and β-$(Al_xGa_{1-x})_2O_3$ devices [32-37]. Among these, $Al_2O_3$ has been extensively investigated and utilized for device applications due to its good compatibility with β-$Ga_2O_3$. Excellent transistor performance has been demonstrated in β-$Ga_2O_3$-based lateral and vertical FET structures, utilizing atomic layer deposited (ALD) $Al_2O_3$ as the gate dielectric, exhibiting a high figure of merit alongside high breakdown field strength [10,38-40]. A recent study focusing on metal-oxide-semiconductor capacitors, fabricated through plasma-assisted deposition of $Al_2O_3$ on β-$Ga_2O_3$, revealed high reverse breakdown electric fields of up to 5.3 MV/cm in the β-$Ga_2O_3$, accompanied by minimal hysteresis in capacitance-voltage and low leakage current [41]. However, in most reported studies, $Al_2O_3$ deposition primarily relies on *ex-situ* techniques such as ALD, involving the transfer of the β-$Ga_2O_3$ substrate or epilayer to a different reactor for dielectric deposition. Such *ex-situ* deposition of $Al_2O_3$ dielectrics may potentially lead to interface contamination due to the exposure of the surface of β-$Ga_2O_3$ epi-films to the ambient environment during loading samples into different chambers for gate dielectric deposition. In contrast, *in-situ* deposition of the gate dielectric offers a viable solution of this challenge by avoiding exposure of the epi-layer to the ambient atmosphere, thereby circumventing potential surface or interface contamination.

Previous research has indicated that the MOCVD *in-situ* growth of amorphous $Al_2O_3$ dielectric on GaN results in lower trap densities ($7.4 \times 10^{11}$ cm$^{-2}$ eV$^{-1}$) and minimal hysteresis compared to *ex-situ* ALD $Al_2O_3$ ($1.6 \times 10^{12}$ cm$^{-2}$ eV$^{-1}$) [42]. Recent investigations have also highlighted the significant impact of *in-situ* MOCVD deposition method as compared to *ex-situ* ALD on both the quality of $Al_2O_3$ dielectrics and the band offsets at dielectric/semiconductor interfaces [43]. The advantages of *in-situ* MOCVD $Al_2O_3$ deposition on β-$Ga_2O_3$ have also been reported with reduced interface traps [44, 45], suggesting high potential for depositing high-quality gate dielectrics using MOCVD. The high-temperature deposition of MOCVD with higher deposition rates can also enhance the quality of bulk dielectric materials with reduced trap density at the interface. The quality of the gate dielectric, specifically its interface trap density ($D_{it}$), fixed charge, thickness, and crystallinity, plays a critical role not only in breakdown behavior but also in determining surface channel mobility in $Ga_2O_3$ MOSFETs. Poor dielectric interfaces can lead to charge trapping and Coulomb scattering, resulting in channel depletion and degraded mobility. Studies have shown that improvements in interface quality can reduce voltage hysteresis and enhance breakdown performance, which are commonly associated with lower $D_{it}$ and fixed charge levels [46-48]. In addition to interface quality, oxide thickness is a key parameter influencing electrostatics and carrier transport. While thinner oxides provide stronger gate control and enhanced electrostatic coupling to the channel, they may increase tunneling leakage currents. Conversely, thicker oxides help suppress leakage but reduce gate-channel coupling, potentially degrading mobility [49]. Therefore, achieving a proper balance between gate control, leakage performance, and transport characteristics is essential for optimizing $Ga_2O_3$-based MOS devices. Furthermore, the crystallinity of the dielectric layer can also be a crucial factor that may affect both interface trap behavior and fixed charge distribution. Increased crystallinity can potentially be correlated with improved dielectric performance in terms of breakdown and stability, and it may also contribute to more stable and efficient carrier transport by reducing structural disorder at the dielectric-semiconductor interface. While the *in-situ* MOCVD growth of $Al_2O_3$ on β-$Ga_2O_3$ has shown promise in addressing some of these challenges, a comprehensive investigation into the interface quality, crystallinity, and electrical performance of *in-situ* MOCVD-deposited $Al_2O_3$ on both β-$Ga_2O_3$ and β-$(Al_xGa_{1-x})_2O_3$ remains limited. A deeper understanding of how growth parameters and alloy composition in β-$(Al_xGa_{1-x})_2O_3$ influence these properties is essential for advancing $Ga_2O_3$-based devices for high-power and RF applications. In this study, we have performed the *in-situ* MOCVD growth of $Al_2O_3$ on both β-$Ga_2O_3$ and β-$(Al_xGa_{1-x})_2O_3$, followed by the electrical characterization of metal/$Al_2O_3$/β-$Ga_2O_3$ (β-$(Al_xGa_{1-x})_2O_3$) vertical MOSCAPs. We have investigated the electrical and structural quality of *in-situ* MOCVD $Al_2O_3$ through capacitance-voltage (C-V) measurements, forward/reverse breakdown measurements, and scanning transmission electron microscopy (STEM) techniques to understand the impact of $Al_2O_3$ deposition temperature and variation of Al compositions of underlying β-$(Al_xGa_{1-x})_2O_3$ films on the crystallinity and current-blocking capabilities of the MOS structure.

## II. Experimental Details

The *in-situ* MOCVD $Al_2O_3$ dielectric layers were deposited on MOCVD grown Si doped β-$Ga_2O_3$ (400 nm thick) and β-$(Al_xGa_{1-x})_2O_3$ (200 nm thick) films on top of (010) oriented Sn doped β-$Ga_2O_3$ conductive substrates (purchased from Novel Crystal Technology) using Agnitron Agilis MOCVD reactor. Triethylgallium (TEGa) and Trimethylaluminum (TMAl) were used as Ga and Al precursors, respectively. Pure $O_2$ and Argon (Ar) were used as $O_2$ precursor and carrier gas, respectively. The $O_2$ flow rate was set at 500 SCCM for all growths. Silicon dopants in the β-$Ga_2O_3$ and β-$(Al_xGa_{1-x})_2O_3$ thin films were introduced into the chamber by flowing diluted silane ($SiH_4$). β-$Ga_2O_3$ and β-$(Al_xGa_{1-x})_2O_3$ growth temperatures were set at 880 °C. The chamber pressures of 60 and 20 Torr were used to grow β-$Ga_2O_3$ and β-$(Al_xGa_{1-x})_2O_3$ thin films, respectively. All substrates were first cleaned *ex-situ* by using solvents (acetone, isopropanol (IPA) and DI water) and then *in-situ* by high temperature (920 °C) annealing for 5 min before the growths initiate. On top of the 400 nm thick β-$Ga_2O_3$ and 200 nm thick β-$(Al_xGa_{1-x})_2O_3$ films, a 25-40 nm thick $Al_2O_3$ dielectric layer was deposited *in-situ* at a growth temperature of 650 and 900 °C (on β-$Ga_2O_3$) and 800 °C (on β-$(Al_xGa_{1-x})_2O_3$) at a chamber pressure of 20 Torr using TMAl as Al precursor. The details of the MOCVD growth parameters for β-$Ga_2O_3$, β-$(Al_xGa_{1-x})_2O_3$ and $Al_2O_3$ dielectrics can be found in our previous reports [2, 18, 19, 43]. To fabricate the MOSCAPs, a Ti(30 nm)/Au (100 nm) ohmic contact stack was deposited by e-beam evaporation on the back side of the samples. The contacts were then annealed at 470 °C using rapid thermal annealing in an $N_2$ ambient for 1 min. Subsequently, circular contacts were patterned using optical lithography, and Ni/Au (20/150 nm) was deposited via e-beam evaporation as the anode metal stack, followed by metal liftoff. The MOS capacitors were first subjected to capacitance-voltage (C-V) characterization and subsequently to current-voltage (I-V) measurements using a Keysight B1500 semiconductor device parameter analyzer. The C-V measurements were conducted at 100 kHz. An aberration-corrected Thermo Fisher Scientific Themis-Z scanning transmission electron microscopy was used to obtain high angle annular dark field (HAADF) STEM images and EDX spectral mapping. Film thicknesses were determined using STEM-EDX elemental mapping profiles and cross-sectional high-resolution STEM-HAADF imaging.

## III. Results and Discussions

Figures 1(a) and 1(b) show the schematic structures of the MOSCAPs, which were grown using in-situ MOCVD deposited 40 nm thick $Al_2O_3$ dielectrics grown at temperatures of 650 °C and 900 °C, respectively. The C-V profiles for both MOSCAPs, as shown in Figure 1(c), reveal a strong influence of $Al_2O_3$ deposition temperature on the voltage hysteresis. Notably, the higher $Al_2O_3$ deposition temperature (900 °C) resulted in a significantly lower hysteresis value of ~0.3 V, whereas the lower deposition temperature (650 °C) led to a higher voltage hysteresis of ~3.44 V at a frequency of 100 kHz. The dielectric constant ($\varepsilon_r$) of the $Al_2O_3$ layers was extracted from the saturated accumulation capacitance, yielding values of approximately 7.79 and 8.22 for the substrate temperatures of 650 °C and 900 °C, respectively. Additionally, the carrier concentration depth profiles ($N_d$-$N_a$) derived from the C-V curves for both MOS structures, as shown in Figure 1(d), revealed net carrier concentrations of $9.50 \times 10^{17}$ cm$^{-3}$ to $8.60 \times 10^{17}$ cm$^{-3}$ in β-$Ga_2O_3$ layers.

The reverse and forward breakdown characteristics of the MOSCAPs were also evaluated. Figure 2(a) shows the reverse breakdown voltages for the MOSCAPs with $Al_2O_3$ deposited at 650

°C and 900 °C. The lower deposition temperature of 650 °C resulted in a reduced reverse breakdown voltage of approximately 38.8 V, while the higher deposition temperature of 900 °C demonstrated improved breakdown voltages around 74.5 V. Under reverse bias, a portion of the voltage drop occurred across the thin $Al_2O_3$ film, with the remainder dropping in the β-$Ga_2O_3$ epitaxial layer. Considering this voltage distribution, the parallel-plate electric field in the β-$Ga_2O_3$ layer at breakdown was estimated to be $E_{Br,\ Ga2O3}$ = 2.87 MV/cm and 4.11 MV/cm for MOSCAPs with $Al_2O_3$ deposited at 650 °C and 900 °C, respectively. Additionally, the reverse breakdown field contribution in $Al_2O_3$ dielectric layer increased from $E_{Br,\ Al2O3}$ = 3.69 MV/cm to 5.01 MV/cm as the deposition temperature increases from 650 °C to 900 °C, respectively. The breakdown was destructive in both samples, occurring at the anode metal edge. The reverse I-V characteristics demonstrated excellent low leakage currents up until hard breakdown. The forward I-V characteristics of the MOSCAPs were also analyzed. The forward breakdown voltages were measured to be 29 V and 22.5 V for MOSCAPs with $Al_2O_3$ deposited at 650 °C and 900 °C, respectively, as shown in Figure 2(b). Detailed calculations of the electric field contributions in both the $Al_2O_3$ and β-$Ga_2O_3$ layers under reverse bias breakdown conditions, as well as the breakdown field in the $Al_2O_3$ layers under forward biasing, are provided in the supplementary materials. The corresponding parallel-plate forward breakdown fields at $Al_2O_3$ deposition temperatures of 650 °C and 900 °C were estimated to be 7.25 MV/cm and 5.62 MV/cm, respectively, indicating a higher forward breakdown field for the lower deposition temperature, consistent with the previous reports on $O_2$-plasma-assisted deposition of $Al_2O_3$ dielectrics on (001) β-$Ga_2O_3$ [41]. This trend can be attributed to the dependence of the crystallinity of the dielectric layers on the deposition temperature, as discussed in the subsequent paragraph using high-resolution STEM imaging and EDX elemental mapping.

Figure 3 provides high-resolution STEM imaging and EDX elemental mapping of the $Al_2O_3$ dielectric deposited at 650 °C on β-$Ga_2O_3$. The cross-sectional high-angle annular dark field (HAADF) STEM images confirmed the $Al_2O_3$ layer thickness to be 40 nm. Low-angle annular dark field (LAADF) STEM revealed a ~5 nm thick crystalline region at the $Al_2O_3$/β-$Ga_2O_3$ interface, while the upper $Al_2O_3$ layer remained amorphous, as shown in Figure 3(c). The atomic-resolution STEM imaging confirmed the crystalline layer at the interface, and STEM-EDS spectroscopy indicated uneven interfaces between $Al_2O_3$ and the (010)-oriented β-$Ga_2O_3$, as shown in Figure 3(d). Elemental mapping of Ga, Al, and O, along with the quantitative elemental profiles in Figure 3(e), demonstrated excellent stoichiometry for both the in-situ MOCVD-grown $Al_2O_3$ dielectrics and β-$Ga_2O_3$ layers. The MOSCAPs with $Al_2O_3$ deposited at 900 °C were also analyzed using STEM imaging, as shown in Figure 4. The cross-sectional STEM image in Figure 4(a) revealed a less pronounced wavy interface compared to the 650 °C sample. LAADF imaging (Figure 4(b)) highlighted significantly higher crystallinity in the dielectric layer deposited at 900 °C as compared to the $Al_2O_3$ deposited at 650 °C. Atomic-resolution STEM imaging, as shown in Figure 4(c), revealed crystalline zones within the $Al_2O_3$ layer, with the atomic arrangement closely matching monoclinic θ-$Al_2O_3$, as depicted in the inset. STEM-EDX elemental mapping (Figures 4(d) and 4(e)) also confirmed excellent compositional homogeneity and negligible interdiffusion of Al and Ga at the $Al_2O_3$/β-$Ga_2O_3$ interface.

As shown in Figure 2(b), the forward breakdown fields of the MOSCAPs with $Al_2O_3$ dielectrics deposited at 900 °C were lower than those of the samples deposited at 650 °C. This behavior can be attributed to the increased crystallinity of the dielectric layer at the higher deposition temperature. STEM imaging, particularly the atomic-resolution images in Figure 4(c), reveals that the $Al_2O_3$ layer deposited at 900 °C contains larger crystalline zones and fewer amorphous regions compared to the 650 °C sample (Figure 3(b)). While this enhanced crystallinity improves structural order, it also introduces more grain boundaries, which are known to facilitate higher leakage currents, thereby reducing the forward breakdown field. Under reverse bias, the leakage current from the metal to the semiconductor is blocked by both the $Al_2O_3$ layer and the voltage drop across the depletion region in $Ga_2O_3$. The MOSCAPs with $Al_2O_3$ deposited at 650 °C exhibited lower reverse breakdown voltages compared to those deposited at 900 °C. This reduced breakdown voltage can potentially be related to a higher concentration of negative fixed charges in the dielectric layer, as indicated by the increased voltage hysteresis in Figure 1(c). Notably, such an increase in negative fixed charges has been previously observed at higher deposition temperatures [41]. The elevated fixed charges lead to an additional voltage drop within the dielectric, increasing the built-in electric field and resulting in earlier breakdown under reverse bias conditions.

Expanding upon the study of $Al_2O_3$/β-$Ga_2O_3$ MOSCAPs, the electrical and structural characteristics of $Al_2O_3$/β-$(Al_xGa_{1-x})_2O_3$ MOS structures with varying Al compositions were also investigated. Figures 5(a) and 5(b) illustrate the schematics of MOSCAPs utilizing in-situ MOCVD-grown $Al_2O_3$ dielectrics, all deposited at the same growth temperature of 800 °C, on β-$(Al_xGa_{1-x})_2O_3$ thin films with Al compositions of x = 5.5% and x = 9.2%, respectively. The C-V curves for both samples, shown in Figure 5(c), reveal no significant influence of Al composition in the β-$(Al_xGa_{1-x})_2O_3$ layers on voltage hysteresis (~1.35V). The net carrier concentration profiles ($N_d$ - $N_a$) derived from these C-V curves as shown in Figure 5(d) indicates a decrease in carrier concentration in the β-$(Al_xGa_{1-x})_2O_3$ layers from 7.87 x $10^{17}$ to 6.19 x $10^{17}$ $cm^{-3}$ as the Al composition increases from 5.5% to 9.2%. This trend indicates that higher Al composition correlates with a reduction in carrier concentration, consistent with prior observations [19].

Additionally, the dielectric constant of the $Al_2O_3$ layer, determined from the saturated accumulation capacitance, is ~10.58 and ~9.26 for the MOSCAPs with Al compositions of x = 5.5% and x = 9.2%, respectively.

The reverse and forward I-V characteristics of the $Al_2O_3$/β-$(Al_xGa_{1-x})_2O_3$ MOSCAPs were evaluated to examine the influence of Al composition on the breakdown behavior of the $Al_2O_3$ dielectrics, as presented in Figure 6. The reverse breakdown voltages, shown in Figure 6(a), reveal that the MOSCAP with x = 5.5% exhibited a lower breakdown voltage of 29.7 V, while the MOSCAP with x = 9.2% demonstrated an improved breakdown voltage of 45.4 V. Under reverse bias, the voltage drop was distributed between the thin $Al_2O_3$ dielectric layer and the underlying β-$(Al_xGa_{1-x})_2O_3$ epitaxial layer. The corresponding parallel-plate electric fields at reverse breakdown in the β-$(Al_xGa_{1-x})_2O_3$ layers ($E_{Br, (AlxGa1-x)2O3}$) were estimated to be 2.55 MV/cm for x = 5.5% and 2.90 MV/cm for x = 9.2%, indicating an increase in the breakdown field with higher Al composition. Similarly, the reverse breakdown electric field within the $Al_2O_3$ dielectric also improved, increasing from 2.41 MV/cm to 3.13 MV/cm as the Al composition in the β-$(Al_xGa_{1-x})_2O_3$ layers increased from x = 5.5% to x = 9.2%. The forward I-V characteristics, presented in Figure 6(b), also showed a rise in forward breakdown fields, with the breakdown fields in the $Al_2O_3$ dielectric ($E_{Br\ Al2O3}$) increasing from 5.0 MV/cm to 5.4 MV/cm as the Al composition in the underlying β-$(Al_xGa_{1-x})_2O_3$ layer increased, indicating improved breakdown performance of the MOSCAPs for β-$(Al_xGa_{1-x})_2O_3$ layer with higher Al content.

High-resolution STEM imaging and STEM-EDX mapping was also performed on the $Al_2O_3$/β-$(Al_xGa_{1-x})_2O_3$ MOS structures to understand the impact of Al incorporation in bulk and interfacial qualities of the $Al_2O_3$ dielectrics, as shown in Figures 7 and 8 for x = 5.5% and 9.2%, respectively. The cross-sectional STEM images of (010) β-$(Al_xGa_{1-x})_2O_3$ film with 5.5% and 9.2%

Al compositions, as shown in Figures 7(a)-(b) and 8(a)-(b), respectively, reveal undisturbed monoclinic β-phase structures in β-$(Al_xGa_{1-x})_2O_3$ without any phase transformation, domain rotation or visible extended defects. The cross-sectional images also show the $Al_2O_3$ dielectric layer with approximately 28 nm thickness for x = 5.5% and 25 nm thickness for x = 9.2%, deposited on β-$(Al_xGa_{1-x})_2O_3$ epi-layer, revealing sharp interfaces. However, a crystalline interlayer between $Al_2O_3$ and β-$(Al_xGa_{1-x})_2O_3$ is observed as indicated by the bright contrast at the $Al_2O_3$/β-$(Al_xGa_{1-x})_2O_3$ interface region in the cross-sectional STEM images in Figure 7(b) and 8(c) for both MOS structures with x = 5.5% and 9.2%, respectively. The high-quality interface between (010) β-$Ga_2O_3$ substrate (bright) and MOCVD grown β-$(Al_xGa_{1-x})_2O_3$ epi-films (dark) is indicated by the sharp contrasts as observed in the STEM images in Figure 7(c) and 8(d). STEM-EDX mapping as shown in Figures 7 (d)-(e) and 8(e)-(f) was conducted on the corresponding $Al_2O_3$/β-$(Al_xGa_{1-x})_2O_3$ MOS structures with x = 5.5% and 9.2%, respectively, to evaluate the compositional homogeneity, stoichiometry and Al compositions. The Ga (green) and Al (blue) EDX color maps in Figures 7(d) and 8(e) indicate no compositional segregations in the β-$(Al_xGa_{1-x})_2O_3$ epitaxial layer. The average Al composition of 5.5% and 9.2% are confirmed from the STEM-EDX elemental mapping profile in Figure 7(e) and 8(f). The STEM-EDX elemental mapping profiles also reveal excellent $Al_2O_3$ stoichiometry for both samples, confirming high quality deposition of in-situ MOCVD $Al_2O_3$ dielectrics on β-$(Al_xGa_{1-x})_2O_3$ films.

## IV. Conclusion

This study comprehensively investigated the influence of $Al_2O_3$ deposition temperature and Al composition in β-$(Al_xGa_{1-x})_2O_3$ films on the electrical and structural properties of in-situ MOCVD $Al_2O_3$/β-$Ga_2O_3$ and $Al_2O_3$/β-$(Al_xGa_{1-x})_2O_3$ MOSCAPs. For $Al_2O_3$/β-$Ga_2O_3$ MOSCAPs, increasing the $Al_2O_3$ deposition temperature from 650 °C to 900 °C significantly reduced voltage hysteresis, indicating improved dielectric quality. This improvement was accompanied by an increase in reverse breakdown voltage from 38.8 V to 74.5 V, with breakdown fields in both the β-$Ga_2O_3$ layer and the $Al_2O_3$ dielectric also improving. However, forward breakdown fields were higher for the 650 °C samples compared to those at 900 °C, a trend attributed to the increased crystallinity and grain boundaries in the dielectric layer at the higher deposition temperature, which contribute to leakage currents. Expanding to $Al_2O_3$/β-$(Al_xGa_{1-x})_2O_3$ MOSCAPs, with Al compositions of x = 5.5% and x = 9.2%, a higher Al incorporation in the β-$(Al_xGa_{1-x})_2O_3$ layers reduced the net carrier concentration. This reduction along with the increase of the bandgap energy of β-$(Al_xGa_{1-x})_2O_3$ for higher Al incorporation correlated with an increase in reverse breakdown voltages from 29.7 V to 45.4 V and an improvement in the corresponding breakdown electric fields in both $Al_2O_3$ dielectric and β-$(Al_xGa_{1-x})_2O_3$ layers as the Al composition increased. Forward breakdown fields similarly showed an improvement with higher Al composition. High-resolution STEM and STEM-EDX analysis confirmed sharp, high-quality interfaces, along with excellent stoichiometry in the $Al_2O_3$ dielectric and the β-$Ga_2O_3$ and β-$(Al_xGa_{1-x})_2O_3$ semiconductor layers. These findings highlight the critical role of in-situ MOCVD $Al_2O_3$ deposition temperature and the Al composition of β-$(Al_xGa_{1-x})_2O_3$ in tailoring the dielectric and breakdown properties of MOSCAPs, providing valuable insights for developing high-performance electronic devices using in-situ MOCVD $Al_2O_3$ dielectrics with β-$Ga_2O_3$ and β-$(Al_xGa_{1-x})_2O_3$.

**Supplementary material**

See the supplementary material for detailed calculations of the electric field contributions in both the $Al_2O_3$ and β-$Ga_2O_3$ layers under reverse bias breakdown conditions, as well as the parallel plate electric field in the $Al_2O_3$ layers calculated under forward bias breakdown conditions for both $Al_2O_3$/β-$Ga_2O_3$ and $Al_2O_3$/β-$(Al_xGa_{1-x})_2O_3$ MOSCAPs.

**Acknowledgements**

The authors acknowledge the financial support from the Air Force Office of Scientific Research FA9550-18-1-0479 (AFOSR, Dr. Ali Sayir), the National Science Foundation (Grant No. 2231026) and the Advanced Research Projects Agency-Energy (ARPA-E), U.S. Department of Energy, under Award Number DE-AR0001036.

**Data Availability**

The data that support the findings of this study are available from the corresponding author upon reasonable request.

## Figure Captions

**Figure 1.** Schematics of the $Al_2O_3$/β-$Ga_2O_3$ MOSCAPs with $Al_2O_3$ deposited at (a) 650 °C and (b) 900 °C. (c) C-V characteristics of the MOSCAPs for $Al_2O_3$ deposited at varying temperatures (frequency at 100 kHz). (d) net carrier concentration profile as a function of depth, extracted from the C-V profiles.

**Figure 2.** (a) Reverse and (b) forward J-V characteristics of $Al_2O_3$/β-$Ga_2O_3$ MOSCAPs with $Al_2O_3$ deposited at 650 and 900 °C.

**Figure 3.** High resolution cross-sectional (a, c) HAADF and (b) LAADF- STEM images of $Al_2O_3$/β-$Ga_2O_3$ MOSCAPs with $Al_2O_3$ deposited at 650 °C. (d) Cross-sectional HAADF image with corresponding EDX mapping of Ga, Al, Ni and O atoms. (e) Atomic fraction elemental profile along the yellow arrow in (d).

**Figure 4.** High resolution cross-sectional (a, c) HAADF and (b) LAADF- STEM images of $Al_2O_3$/β-$Ga_2O_3$ MOSCAPs with $Al_2O_3$ deposited at 900 °C. (d) Cross-sectional HAADF image with corresponding EDX mapping of Ga, Al, Ni and O atoms. (e) Atomic fraction elemental profile along the yellow arrow in (d).

**Figure 5.** Schematics of the $Al_2O_3$β-$(Al_xGa_{1-x})_2O_3$ MOSCAPs with (a) x = 5.5% and (b) x = 9.2%. (c) C-V characteristics of the MOSCAPs for $Al_2O_3$ deposited at varying temperature. (d) Net carrier concentration profile as a function of depth, extracted from the C-V profiles.

**Figure 6.** (a) Reverse and (b) forward J-V characteristics of $Al_2O_3$/β-$(Al_xGa_{1-x})_2O_3$ MOSCAPs with x = 5.5% and 9.2%.

**Figure 7.** High resolution cross-sectional (a, c) HAADF and (b) LAADF- STEM images of $Al_2O_3$/β-$(Al_xGa_{1-x})_2O_3$ MOSCAPs with x = 5.5%. The *in-situ* MOCVD $Al_2O_3$ layer was deposited at 800 °C on top of β-$(Al_xGa_{1-x})_2O_3$ (x = 5.5%) layer. (d) Cross-sectional HAADF image with corresponding EDX mapping of Ga, Al, Ni and O atoms. (e) Atomic fraction elemental profile along the yellow arrow in (d).

**Figure 8.** High resolution cross-sectional (a,c,d) HAADF and (b) LAADF- STEM images of $Al_2O_3$/β-$(Al_xGa_{1-x})_2O_3$ MOSCAPs with x = 9.2%. The *in-situ* MOCVD $Al_2O_3$ layer was deposited at 800 °C on top of β-$(Al_xGa_{1-x})_2O_3$ (x = 9.2%) layer. (e) Cross-sectional HAADF image with corresponding EDX mapping of Ga, Al, Ni and O atoms. (f) Atomic fraction elemental profile along the yellow arrow in (d).

**Figure 1**

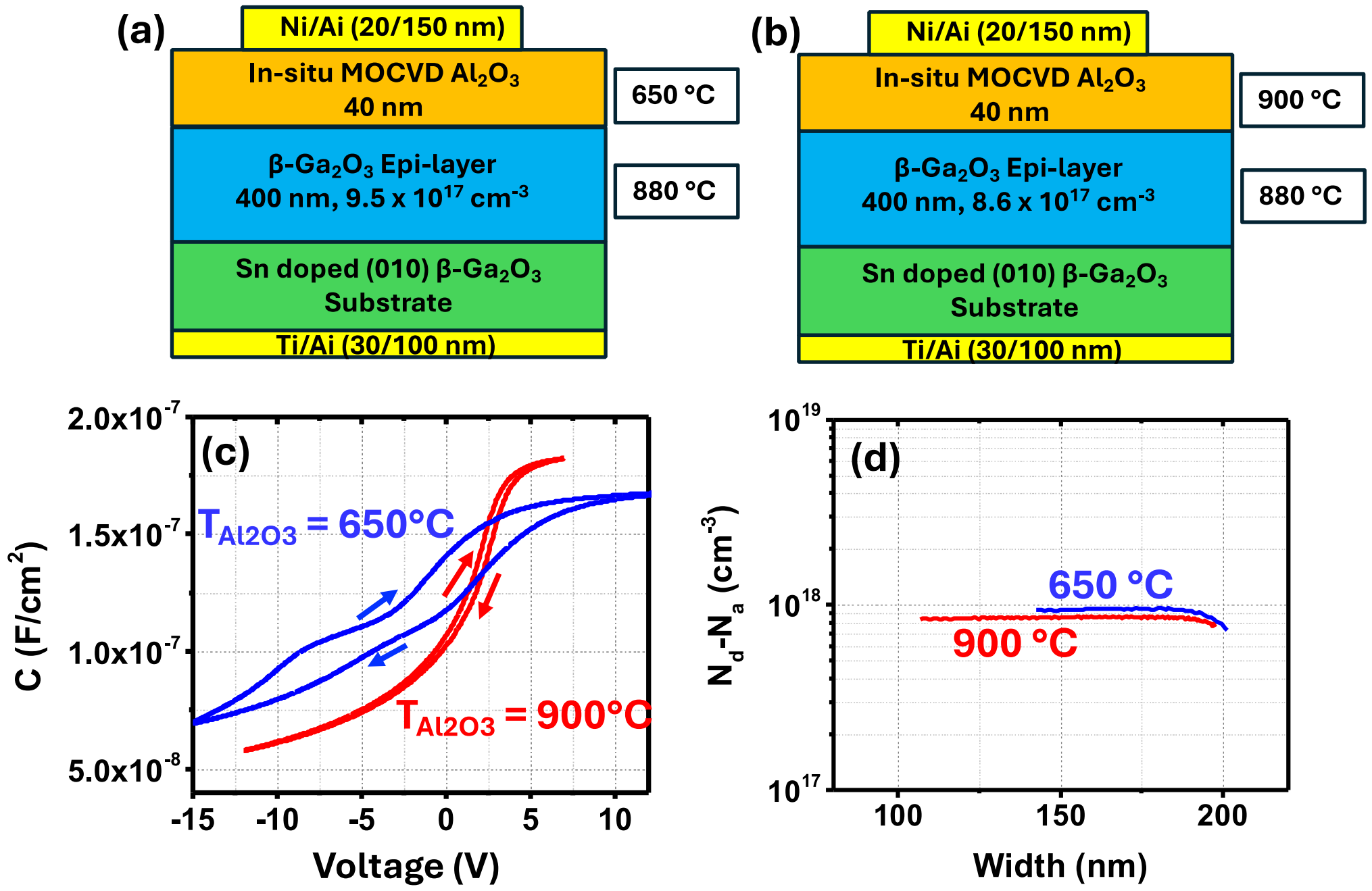

**Figure 1.** Schematics of the $Al_2O_3$/β-$Ga_2O_3$ MOSCAPs with $Al_2O_3$ deposited at (a) 650 °C and (b) 900 °C. (c) C-V characteristics of the MOSCAPs for $Al_2O_3$ deposited at varying temperatures (frequency at 100 kHz). (d) net carrier concentration profile as a function of depth, extracted from the C-V profiles.

**Figure 2**

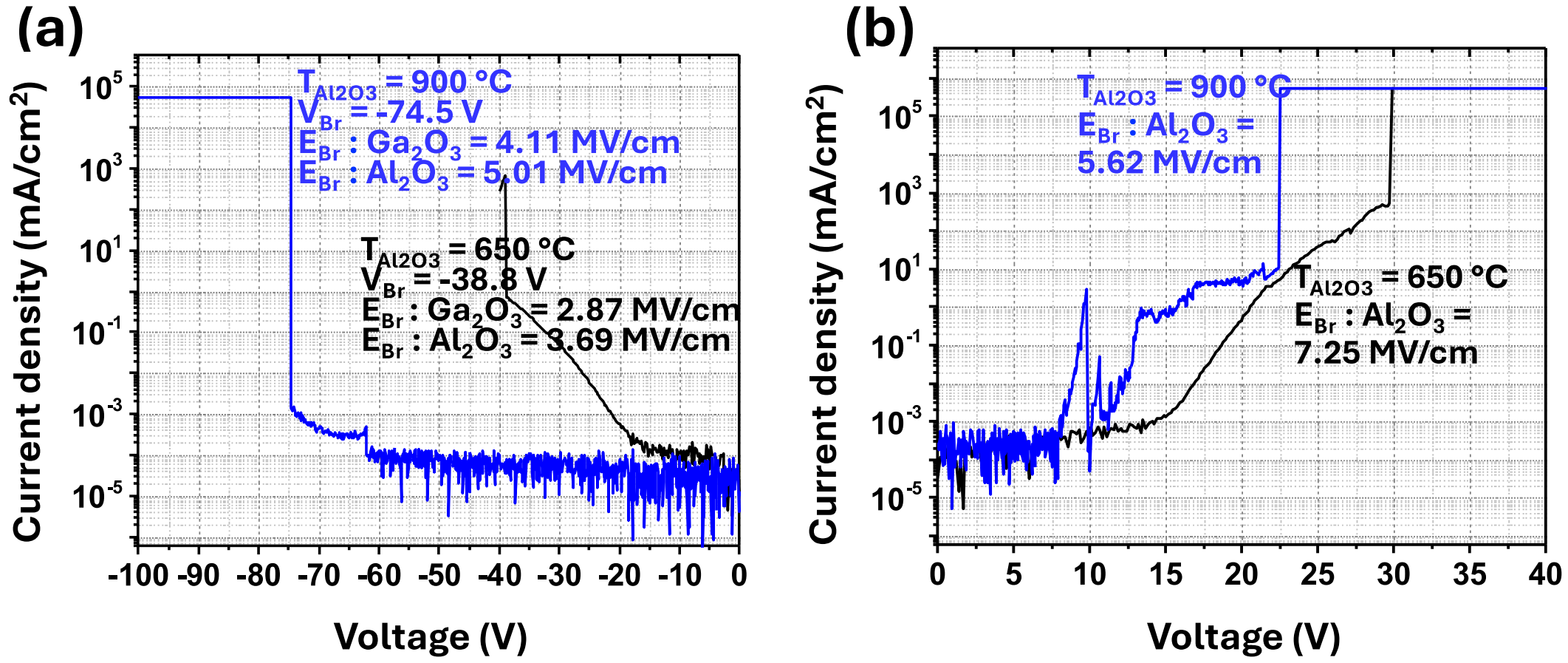

**Figure 2.** (a) Reverse and (b) forward J-V characteristics of $Al_2O_3$/β-$Ga_2O_3$ MOSCAPs with $Al_2O_3$ deposited at 650 and 900 °C.

**Figure 3**

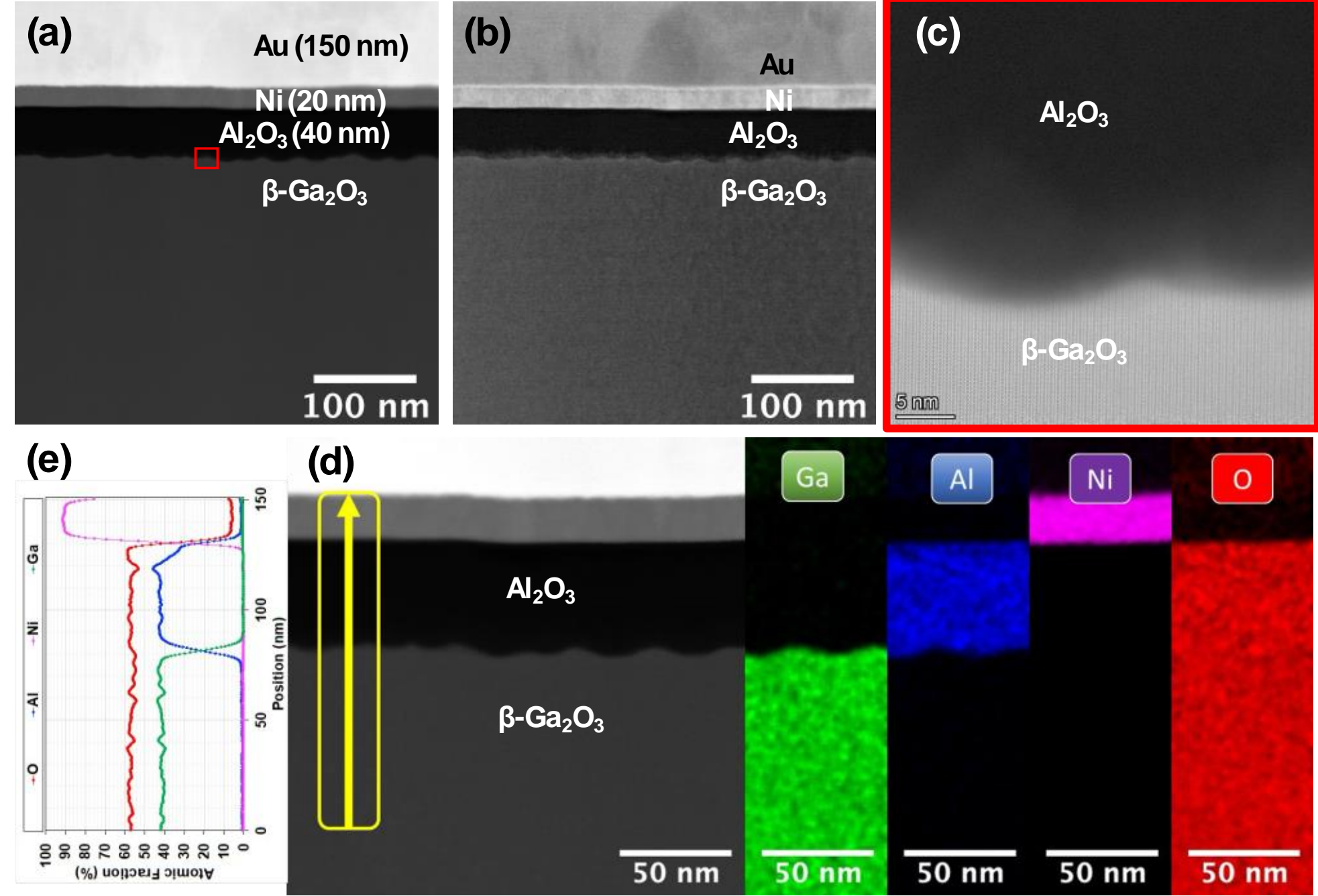

**Figure 3.** High resolution cross-sectional (a, c) HAADF and (b) LAADF- STEM images of $Al_2O_3$/β-$Ga_2O_3$ MOSCAPs with $Al_2O_3$ deposited at 650 °C. (d) Cross-sectional HAADF image with corresponding EDX mapping of Ga, Al, Ni and O atoms. (e) Atomic fraction elemental profile along the yellow arrow in (d).

**Figure 4**

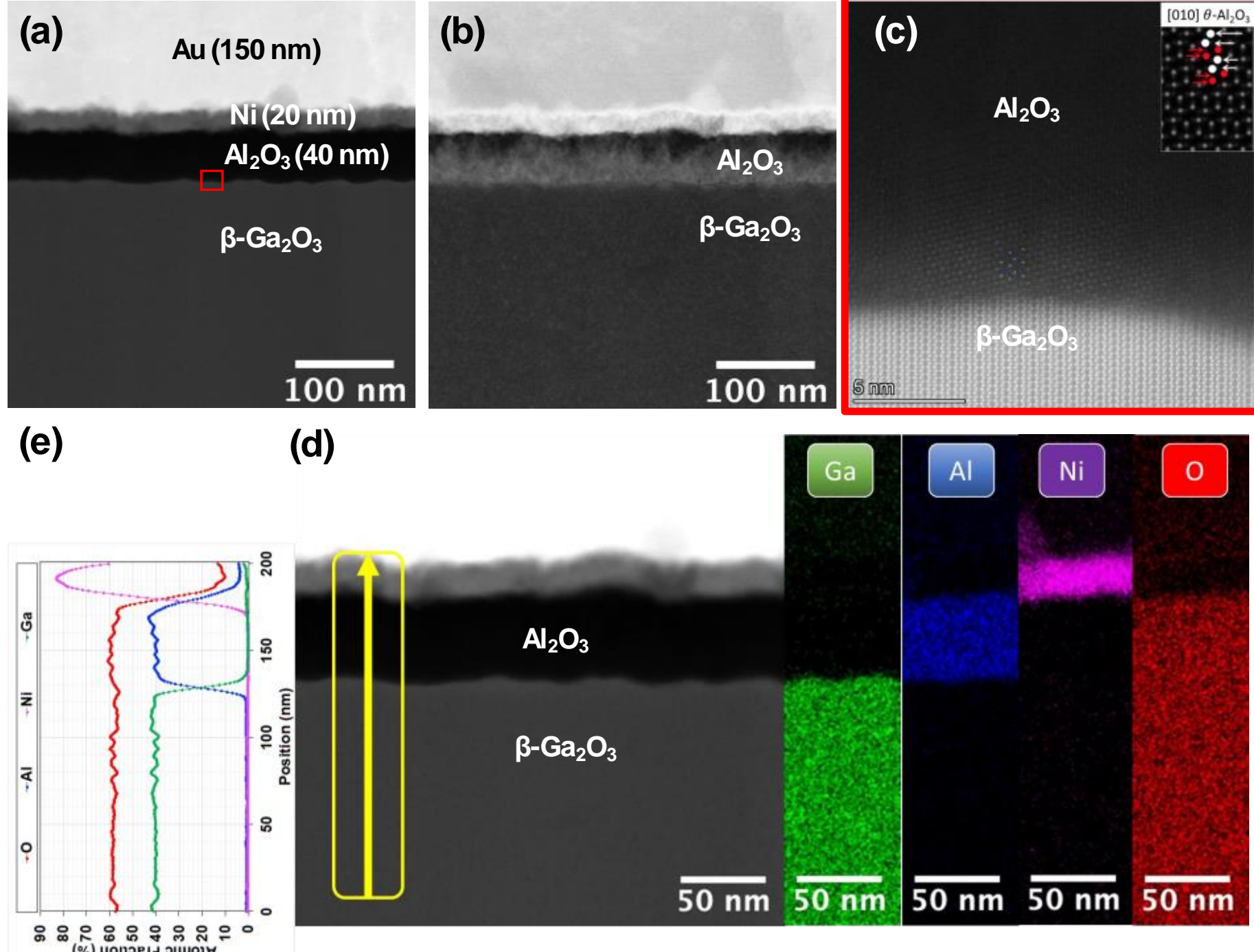

**Figure 4.** High resolution cross-sectional (a, c) HAADF and (b) LAADF- STEM images of $Al_2O_3$/β-$Ga_2O_3$ MOSCAPs with $Al_2O_3$ deposited at 900 °C. (d) Cross-sectional HAADF image with corresponding EDX mapping of Ga, Al, Ni and O atoms. (e) Atomic fraction elemental profile along the yellow arrow in (d).

**Figure 5**

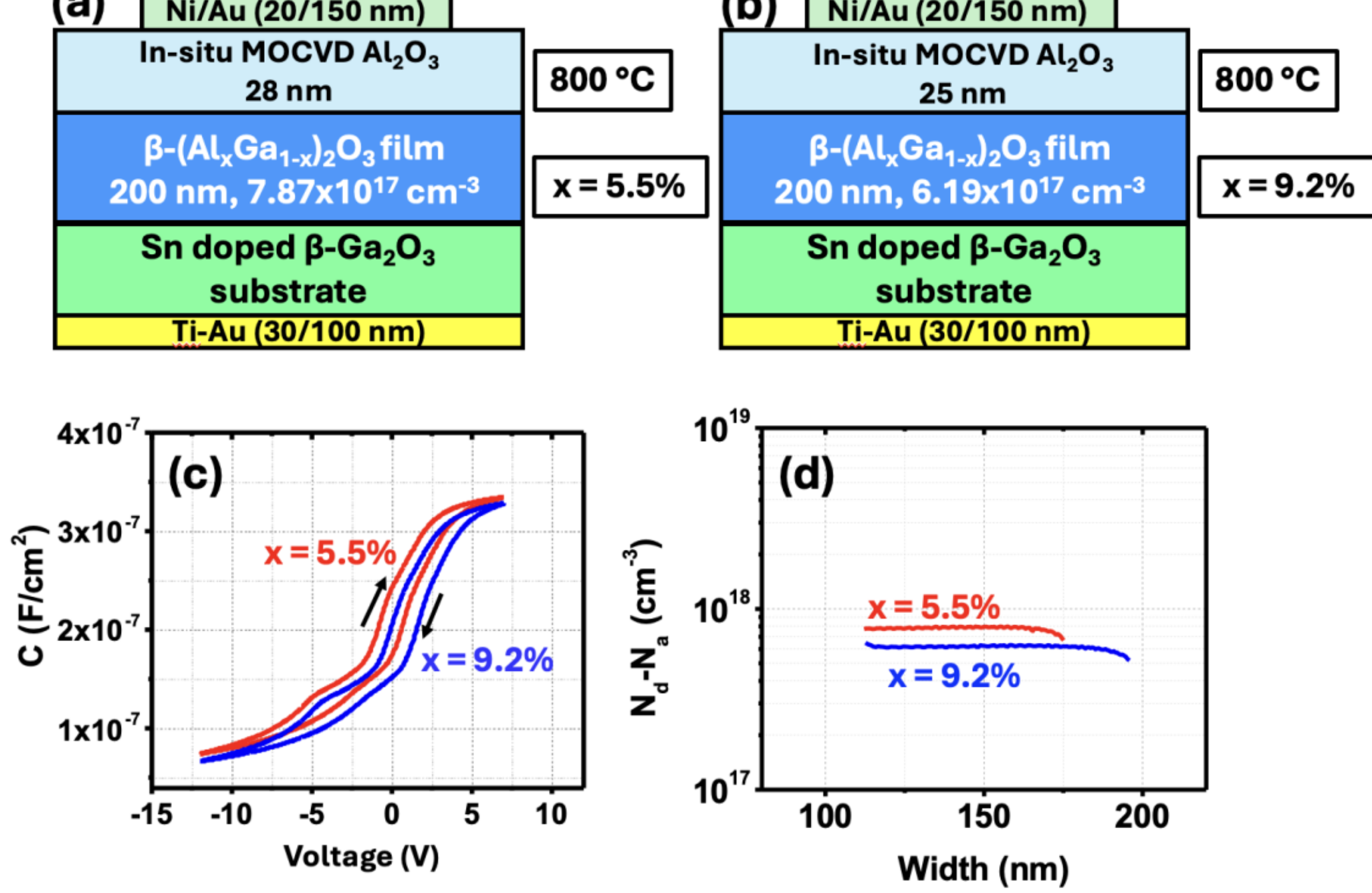

**Figure 5.** Schematics of the $Al_2O_3$ β-$(Al_xGa_{1-x})_2O_3$ MOSCAPs with (a) x = 5.5% and (b) x = 9.2%. (c) C-V characteristics of the MOSCAPs for $Al_2O_3$ deposited at varying temperature. (d) Net carrier concentration profile as a function of depth, extracted from the C-V profiles.

**Figure 6**

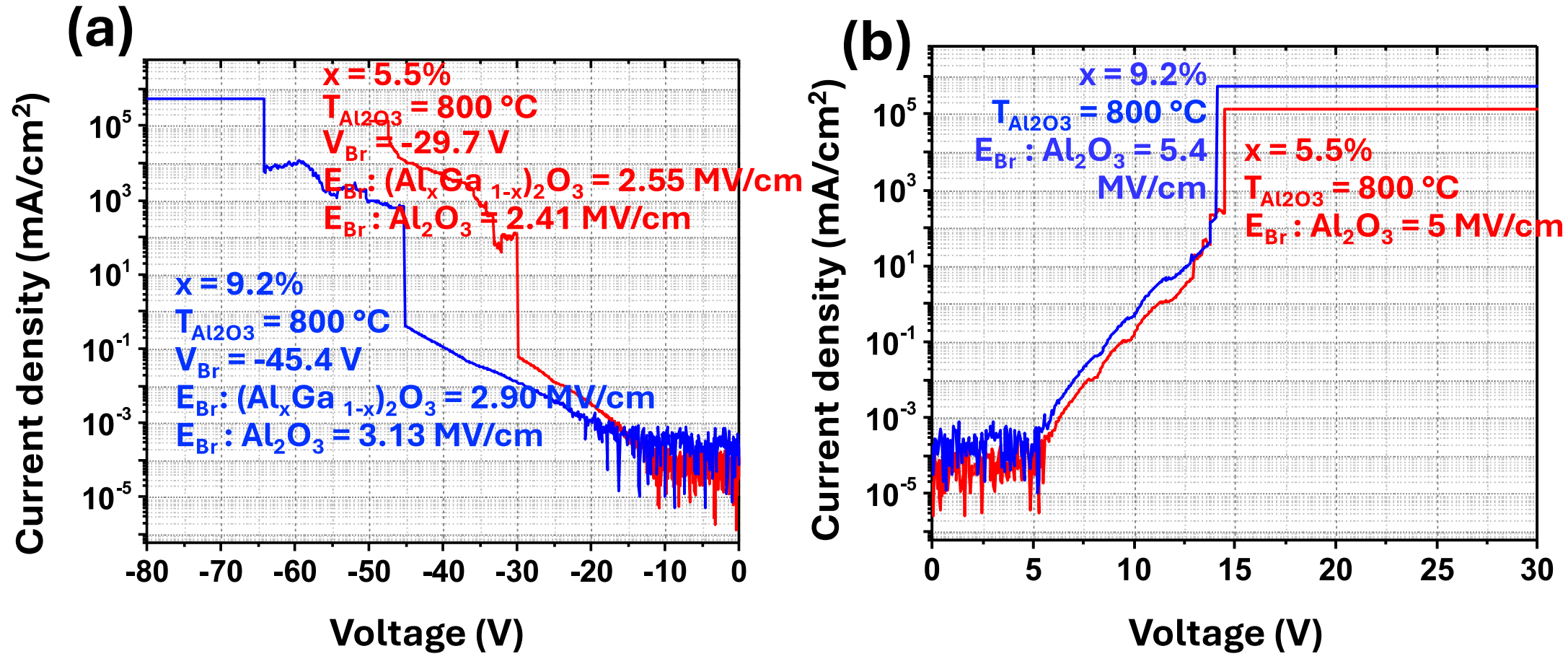

**Figure 6.** (a) Reverse and (b) forward J-V characteristics of $Al_2O_3$/β-$(Al_xGa_{1-x})_2O_3$ MOSCAPs with x = 5.5% and 9.2%.

.

**Figure 7**

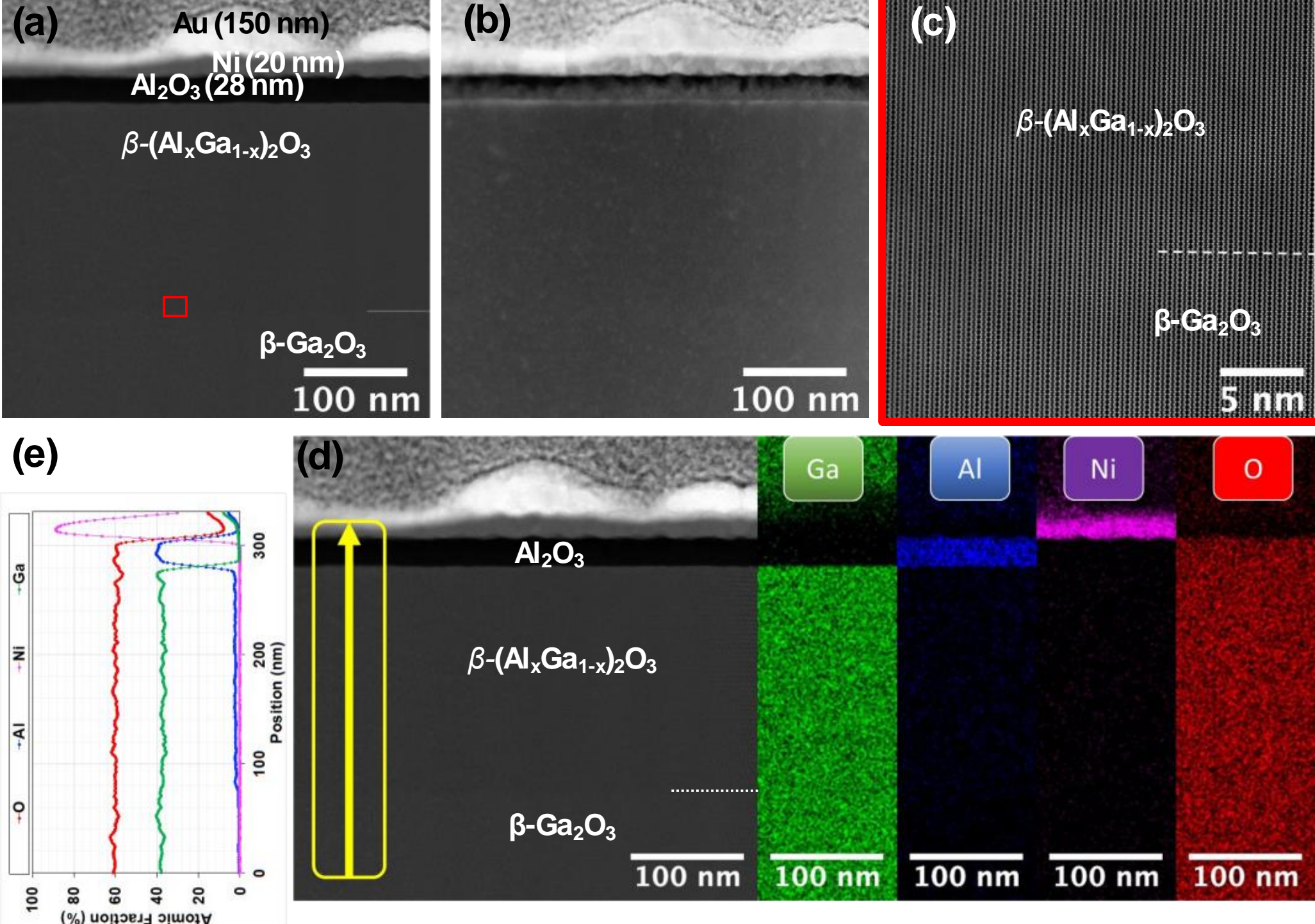

**Figure 7.** High resolution cross-sectional (a, c) HAADF and (b) LAADF- STEM images of $Al_2O_3$/β-$(Al_xGa_{1-x})_2O_3$ MOSCAPs with x = 5.5%. The *in-situ* MOCVD $Al_2O_3$ layer was deposited at 800 °C on top of β-$(Al_xGa_{1-x})_2O_3$ (x = 5.5%) layer. (d) Cross-sectional HAADF image with corresponding EDX mapping of Ga, Al, Ni and O atoms. (e) Atomic fraction elemental profile along the yellow arrow in (d).

## Figure 8

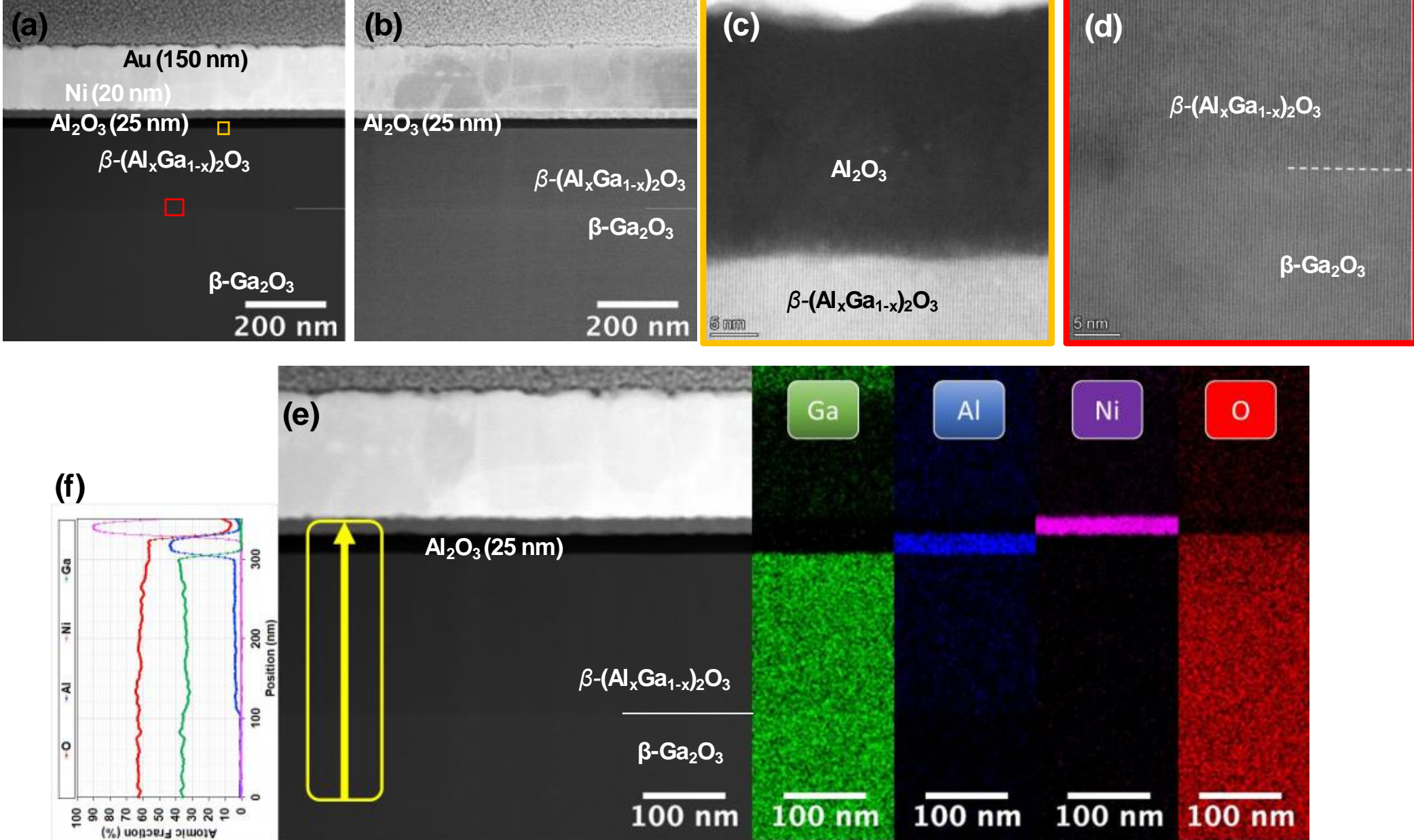

**Figure 8.** High resolution cross-sectional (a,c,d) HAADF and (b) LAADF- STEM images of $Al_2O_3$/β-$(Al_xGa_{1-x})_2O_3$ MOSCAPs with x = 9.2%. The *in-situ* MOCVD $Al_2O_3$ layer was deposited at 800 °C on top of β-$(Al_xGa_{1-x})_2O_3$ (x = 9.2%) layer. (e) Cross-sectional HAADF image with corresponding EDX mapping of Ga, Al, Ni and O atoms. (f) Atomic fraction elemental profile along the yellow arrow in (d).